\documentclass[camera,letterpaper,nomarginnotes,nonarrowgutter]{jpaper}

\usepackage{graphicx}
\usepackage{textcomp}
\usepackage{fancyvrb}
\usepackage{ulem}
\usepackage{fancyhdr}
\usepackage{flushend}
\usepackage{comment}
\usepackage{amsmath}
\usepackage{bbding}
\usepackage[sort,compress]{cite}
\usepackage{balance}
\usepackage{titling}
\usepackage[bookmarks=true,breaklinks=true,letterpaper=true,colorlinks,citecolor=blue,linkcolor=blue,urlcolor=blue]{hyperref}

\title{A Simulation-based Evaluation Framework for\\Inter-VM RowHammer Mitigation Techniques}

\author{}
\author{Hidemasa Kawasaki$\dagger$ \quad Soramichi Akiyama$\dagger$
\vspace{-2mm}
\\\\\emph{Ritsumeikan University}$\dagger$}

\begin{document}
\bstctlcite{IEEEexample:BSTcontrol}
\maketitle

%Enables the camera ready header and footer                                                                                                                                                   

\thispagestyle{plain}
\pagestyle{plain}

\begin{abstract}
Inter-VM RowHammer is an attack that induces a bitflip beyond the boundaries of virtual machines (VMs) to compromise a VM from another,
and some software-based techniques have been proposed to mitigate this attack.
Evaluating these mitigation techniques requires to confirm that they actually mitigate inter-VM RowHammer in low overhead.
A challenge in this evaluation process is that both the mitigation ability and the overhead depend on the underlying hardware whose DRAM address mappings are different from machine to machine. 
This makes comprehensive evaluation prohibitively costly or even implausible as no machine that has a specific DRAM address mapping might be available.
To tackle this challenge, we propose a simulation-based framework to evaluate software-based inter-VM RowHammer mitigation techniques across configurable DRAM address mappings.
We demonstrate how to reproduce existing mitigation techniques on our framework, and show that it can evaluate the mitigation abilities and performance overhead of them with configurable DRAM address mappings.
\end{abstract}

\section{Introduction}
Inter-VM RowHammer attack is a serious concern in multi-tenant cloud environments.
In this attack, an attacker triggers RowHammer~\cite{Kim2014} from their VM to compromise the VMs of other tenants or even the hypervisor.
It poses a significant threat to hypervisor-based isolation,
which is often the strongest security measure employed in cloud environments.

Two software-based inter-VM RowHammer mitigation techniques have been proposed:
Siloz~\cite{Kevin2023} and Citadel~\cite{saxena2024}.
Siloz leverages the observation that the RowHammer effect is confined to DRAM subarrays and isolates VMs by allocating their memory to different subarrays. 
Citadel inserts unused DRAM rows, named guard rows, as buffers between the memory regions allocated to different VMs.

Evaluating software-based inter-VM RowHammer mitigation techniques typically focuses on two key aspects: security and performance overhead. 
Security involves conducting a RowHammer attack within a VM to verify that the mitigation technique successfully prevents it from affecting other VMs or the hypervisor. 
Performance overhead is quantified by measuring application execution time and effective memory bandwidth within the VMs under the mitigation technique.

A challenge in these evaluation methodologies is that they require a comprehensive understanding of the DRAM address mapping of the machines used.
This is because the DRAM address mapping affects the mitigation ability and the performance overhead of software-based mitigation techniques.
%This means that evaluation without knowing the address mapping is irrelevant.
The challenge is significant because DRAM address mappings vary widely across different CPU models and machine configurations,
and even worse, no machine currently available may have a specific DRAM address mapping.

To overcome this challenge, employing a simulator is a viable approach. 
Simulators can potentially model various hardware configurations and DRAM address mappings, enabling broader evaluation.
However, the issue here is that existing simulators~\cite{Thomas2023, Kaustav2023, Luo2023} focus on the hardware side of memory systems.
Therefore, these simulators cannot straightforwardly reproduce inter-VM RowHammer (e.g., they cannot run VMs as-is) or its mitigation techniques.

In this paper, we propose a simulation framework to evaluate inter-VM RowHammer mitigation techniques across diverse DRAM address mappings.
This is achieved by a whole-system simulation including a hypervisor with configurable DRAM address mappings and an interface to reproduce inter-VM RowHammer mitigation techniques.
%This facilitates exploring scenarios corresponding to different hardware platforms, including those not physically available.
bh. Our case study demonstrates that the framework can reproduce two existing inter-VM RowHammer mitigation techniques and evaluate their mitigation effectiveness and performance overhead on various DRAM address mappings.

\section{Background}
\label{background}
\subsection{DRAM Internals}
The physical organization of DRAM is hierarchical. 
This hierarchy consists of channels, ranks, banks, subarrays, rows, and columns.
Channels are at the top level and each channel is connected to one or more ranks.
A rank can be divided into banks, which are logically independent arrays of memory cells.
Memory requests (reads/writes) can be executed concurrently in different banks, giving parallelism to the upper layer. 
A bank itself is partitioned into smaller units called subarrays.
Each subarray contains a two-dimensional array of memory cells, structured as rows and columns.

A {\it DRAM address mapping} translates a physical address to the corresponding {\it DRAM coordinates} (e.g., channel, rank, bank, subarray, row, and column).
For example, a simple address mapping would be to determine the channel using the most significant bits of a physical address,
and then determine the rank by the subsequent bits, etc.

\subsection{Inter-VM RowHammer Attacks}
RowHammer~\cite{Kim2014} is an attack that exploits disturbance errors in DRAM.
An attacker can frequently access the same DRAM row to induce bitflips in physically adjacent rows, which can be allocated to a different user.
The row that is frequently accessed by the attacker is called the aggressor row and the adjacent rows are called the victim rows.
It is well known that a single bitflip can cause OS-level privilege escalation and other serious threats~\cite{project_zero_rowhammer, Zhang2020, Kwong2020, Yoshioka2024}. 

A major category of RowHammer-related attacks is inter-VM RowHammer.
In this attack, an attacker induces bitflips across VM boundaries to compromise other VMs or the hypervisor itself~\cite{Xiao2016, Razavi2016, Chen2025}.
Inter-VM RowHammer poses a serious threat as it breaks VM boundaries which are often the strongest security measures employed by cloud providers.

\subsection{Inter-VM RowHammer Mitigation Techniques}
\label{section:existing_mitigation_techniques}
\begin{figure}
    \centering
    \includegraphics[width=1\linewidth]{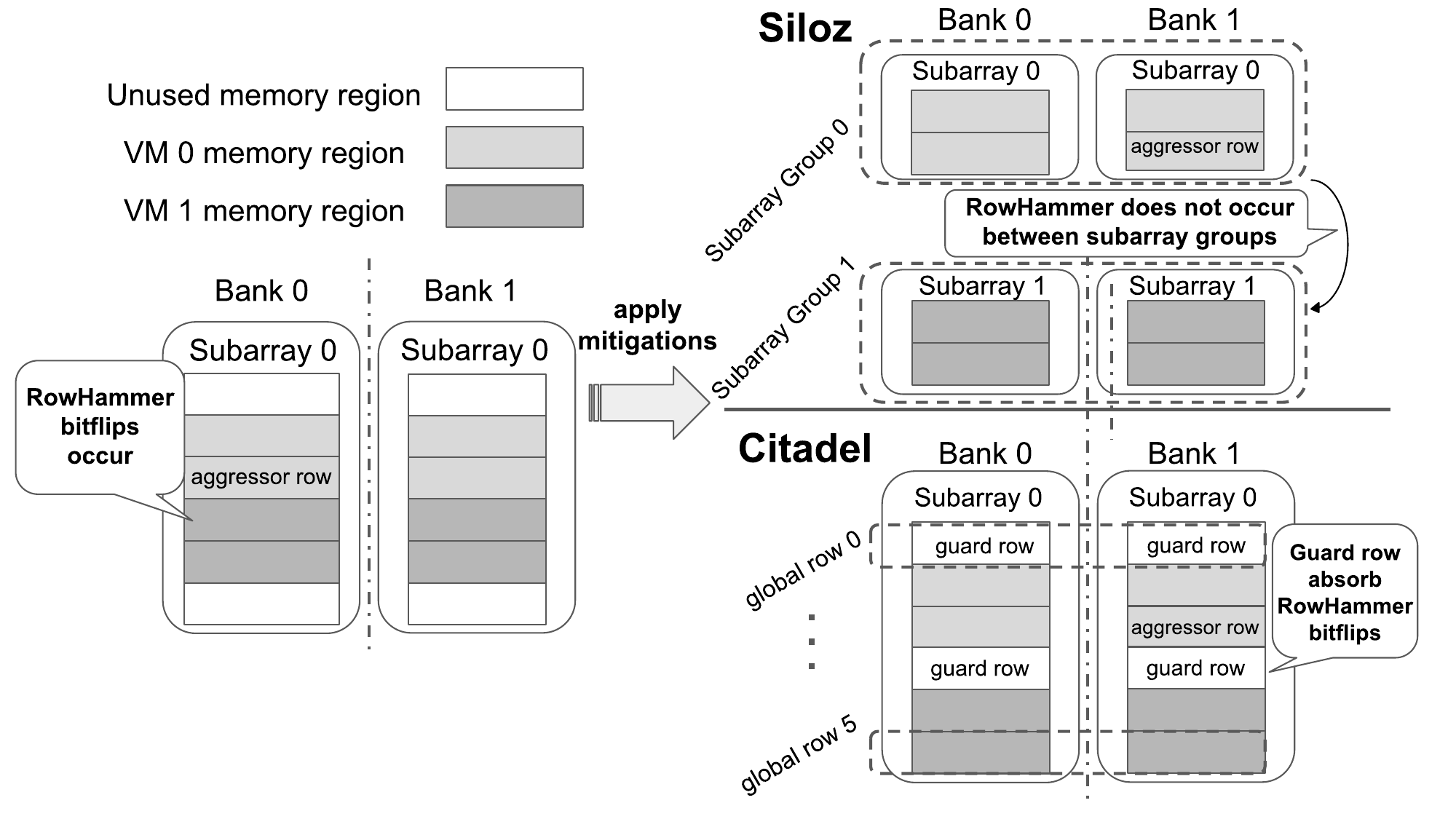}
    \caption{Existing Inter-VM RowHammer Mitigation Techniques}
    \label{fig:a-label}
\end{figure}

To prevent inter-VM RowHammer, two software-based mitigation techniques have been proposed.
Fig.~\ref{fig:a-label} shows how these techniques prevent inter-VM RowHammer.

{\bf Siloz}~\cite{Kevin2023} leverages the fact that the RowHammer effect is confined to DRAM subarrays.
A subarray group is a set of subarrays with the same subarray index across different banks.
Siloz isolates VM memory regions by allocating them to distinct subarray groups.
The memory region allocated to a VM is static and contiguous in the host physical address space for performance reasons (Section 5.4 in~\cite{Kevin2023}).

{\bf Citadel}~\cite{saxena2024} employs guard rows that are unused DRAM rows inserted between the memory regions allocated to different VMs and between VMs and the hypervisor.
These guard rows absorb inter-VM RowHammer-induced bitflips.
Citadel defines a global row as a set of DRAM rows sharing the same row index across all banks
and allocates a set of contiguous global rows to a VM.
A memory region allocated to a VM is not necessarily contiguous in the host physical address space due to the DRAM address mapping. 

\subsection{Challenge in Evaluating Software-based Inter-VM RowHammer Mitigation Techniques}
RowHammer mitigation techniques in general are evaluated in two aspects.
First, we must ensure that they successfully prevent RowHammer attacks in various environments.
Second, we need to quantify the performance overhead to the user-visible system (i.e., the VMs in the inter-VM RowHammer context) incurred by the mitigation technique.

Evaluating inter-VM RowHammer mitigation techniques must account for the DRAM address mapping of the underlying hardware.
This is because they often impose strict constraints on the physical placement of VM memory regions inside DRAM chips (e.g., Siloz contains a VM in a physical subarray group).
This means that evaluating their mitigation ability is irrelevant without knowing the DRAM address mapping.
In addition, such placement requirements can limit the bank-level parallelism or degrade the rowbuffer hit rate.
These could result in lower VM performance compared to a case where VM memory regions can be freely distributed across a DRAM chip for the best performance.

Although important, evaluation with DRAM address mappings taken into account is challenging.
{\bf First}, it is known that DRAM address mappings vary considerably across CPU models and machine configurations~\cite{Pessl2016,Helm2020,Gerlach2024}.
This means that no representative mapping exists and the evaluation must rely on a diverse set of hardware.
{\bf Second}, it is possible that no machine currently available in the world has a specific DRAM address mapping.
In this case, evaluation with this DRAM address mapping on real hardware is merely impossible.
Note that it does not mean that this evaluation is meaningless because a machine in the near future could have that particular DRAM address mapping.

\section{Proposed Framework}
\subsection{Overview}
\label{sec:proposed_framework_overview}

\begin{figure}
    \centering
    \includegraphics[width=1\linewidth]{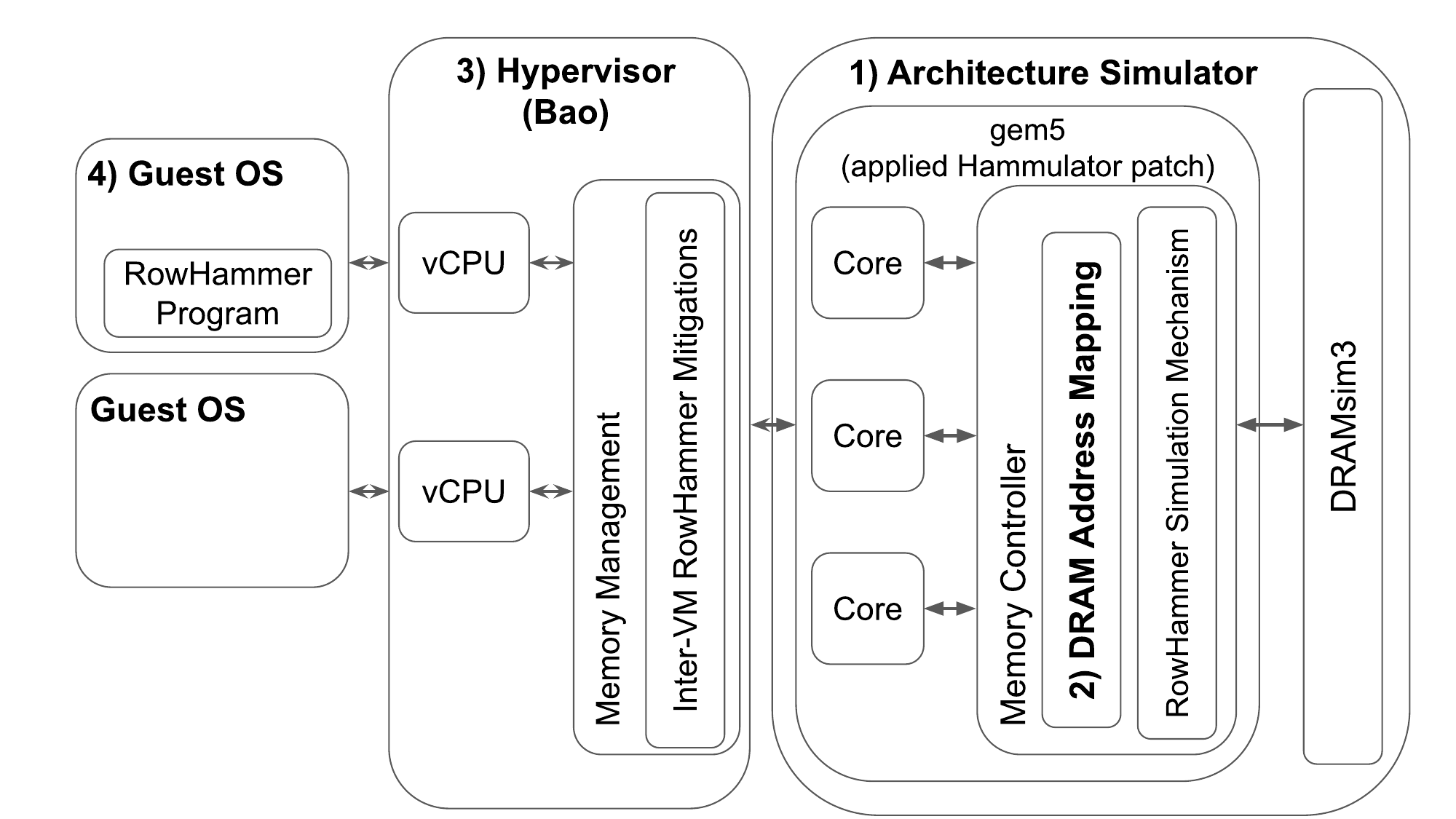}
    \caption{Overview of Our Inter-VM RowHammer Simulation Framework}
    \label{fig:framework-overview}
\end{figure}

We propose a simulation framework that enables evaluating inter-VM RowHammer mitigation techniques under configurable DRAM address mappings.
The {\bf key idea} is to run a hypervisor on an architecture simulator and reproduce inter-VM RowHammer mitigation techniques inside it to avoid the need for reverse engineering DRAM address mappings.
Fig.~\ref{fig:framework-overview} shows an overview of our framework.
It has four {\bf key components} described below:

\begin{enumerate}
    \item An architecture simulator that simulates a CPU and DRAM. 
    It is also capable of simulating RowHammer attacks by considering the frequency of accesses to a particular DRAM row.
    \item A DRAM address mapping function that is configurable by the user. It supports not only simple mappings but also more sophisticated ones with XOR operations.
    \item A lightweight hypervisor running on the architecture simulator.
    It provides an interface to configure its memory management component so that the user can reproduce inter-VM RowHammer mitigation techniques.
    \item Guest VMs running on the hypervisor. The user can execute an attacker program inducing an inter-VM RowHammer attack in a VM and observe if the mitigation technique under evaluation can prevent it.
\end{enumerate}

Our framework enables the evaluation of software-based inter-VM mitigation techniques through four steps:
\begin{enumerate}
    \item \textbf{Define DRAM address mapping}:
    The user specifies a DRAM address mapping that maps host physical addresses to DRAM coordinates.
    This step allows reproducing realistic or hypothetical mappings.
    \item \textbf{Reproduce inter-VM RowHammer mitigation techniques}:
    The user reproduces inter-VM RowHammer mitigation techniques by using the interface provided by the memory management mechanism of the hypervisor.
    %Section~\ref{subsec:Implement_inter-VM_RowHammer_Mitigation_Techniques} provides examples of how specific techniques are implemented within this framework.
    \item \textbf{Induce inter-VM RowHammer}: 
    The user executes a program that conducts a RowHammer attack within one VM.
    We refer to this program as an {\it attacker program}.
    Our framework allows the attacker program to precisely access a particular DRAM row that is adjacent to the memory region of the target VM. 
    This is made possible by the fact that the DRAM address mapping is known and that our hypervisor adopts a straight mapping from guest physical to host physical addresses.
    Due to the latter, we simply call both addresses as {\it physical addresses (PAs)} throughout this paper.
    The framework reports the locations (PAs and DRAM coordinates) of bitflips induced by RowHammer.
    \item \textbf{Evaluate inter-VM RowHammer mitigation techniques}: 
    The user evaluates inter-VM RowHammer mitigation techniques using the reports on bitflip locations generated by our framework and the stats on performance by the underlying architecture simulator.
%    This report allows verification of whether a mitigation implemented in the hypervisor successfully prevents inter-VM bitflips under the defined address mapping.
\end{enumerate}

\subsection{Define DRAM Address Mapping}
\label{sec:dram_mapping_config}
Our framework provides an interface for the user to define a DRAM address mapping based on various translation functions.
These translation functions range from simple contiguous bit selections to complex mappings involving non-contiguous bits and XOR operations.
This capability is important because many platforms have DRAM address mappings with such complexities~\cite{Pessl2016,Helm2020,Gerlach2024}.
The user can specify which PA bits (or XOR of them) represent a particular DRAM coordinate (e.g., banks) through the interface.

Our interface validates a given DRAM address mapping.
This validation ensures that a unique bidirectional conversion exists between the PAs and the DRAM coordinates (i.e., for any PA $A$, there exists one and only one DRAM coordinate $C$ that is mapped from $A$ by the given DRAM address mapping, and vice versa).
The framework employs Gaussian elimination for validation and reports an error if it fails.

\subsection{Reproduce Inter-VM RowHammer Mitigation Techniques}
\label{subsec:Implement_inter-VM_RowHammer_Mitigation_Techniques}
To reproduce inter-VM RowHammer mitigation techniques, we use a hypervisor that satisfies the following conditions:
\begin{enumerate}
    \item It can run multiple VMs on top of it so that both attacker and victim VMs can be simulated simultaneously.
    \item It uses a simple memory management mechanism that is static and straight to facilitate easy reproduction of mitigation techniques.
\end{enumerate}
Here, static means that the memory region for a given VM is allocated at once in its boot-time, and straight means the guest physical addresses are the same as the host physical addresses.

Our framework provides an interface to specify the starting PA and the size of the memory region allocated to a VM.
The user can also create unused regions with this interface.
The interface only supports assigning a single contiguous PA range to a VM.
This is enough for reproducing Siloz because it assumes the same limitation as described in Section~\ref{section:existing_mitigation_techniques}.
For Citadel, it can only reproduce scenarios where the allocated memory region to a single VM is contiguous in terms of DRAM rows within all given channels, ranks, and banks.

\subsection{Induce Inter-VM RowHammer Bitflips}
To induce inter-VM RowHammer within our framework, the user performs the following actions:
\begin{enumerate}
    \item The user launches two VMs whose assigned rows are adjacent to each other. To do this, the user considers the DRAM address mapping they define to allocate memory regions with proper PAs.
    \item The user identifies an aggressor row and calculates its PA.
    An aggressor row must reside in the memory region of the attacker VM and be adjacent to another DRAM row in the memory region of the victim VM.
    \item The user repeatedly accesses the identified PA from the attacker VM.
    We explain how we can access a particular PA from the userspace on Linux in Section~\ref{sec:framework_setup}.
\end{enumerate}

\subsection{Implementation Details}
For the architecture simulator, we adopt gem5~\cite{gem5} and DRAMSim3~\cite{Li2020} with the Hammulator~\cite{Thomas2023} patches applied.
We extend them so that they support complex DRAM address mappings with XORed address bits and the interface for the user to specify mappings.
While other simulators such as Ramulator~\cite{Luo2023} support such address mappings, we chose to extend DRAMsim3 to leverage Hammulator's open-source RowHammer simulation logic. 
This approach allowed us to focus our efforts on the core aspects of our framework rather than re-implementing fundamental capabilities.
We also modify DRAMSim3 to simulate subarray-level isolation; accessing a row in a subarray in our modified version does not affect rows in other subarrays.
We use the full-system mode of gem5 with an O3 (Out-Of-Order) CPU and the ARM ISA.
ARM is the only ISA in gem5 that implements hypervisor-related instructions\footnote{A series of patches that implement the RISC-V Hypervisor extension was merged to gem5 after the acceptance of this paper. \url{https://github.com/gem5/gem5/pull/1387}}.
The use of a specific ISA does not hurt the generality of our framework because its design is ISA-independent.

For the hypervisor, we extend Bao~\cite{Martins2020} because
(1) it satisfies the required conditions in Section~\ref{subsec:Implement_inter-VM_RowHammer_Mitigation_Techniques} and
(2) it runs on ARM Fixed Virtual Platform~\cite{ARM_FVP} that gem5 can readily simulate.
We modify Bao so it receives the configuration on VM memory regions from our framework.
We use Trusted Firmware-A~\cite{TrustedFirmware} and U-Boot~\cite{U-Boot} to boot Bao on gem5.

\section{Case Study}
\label{Evaluation}
As a case study of our framework, we reproduce existing software-based inter-VM RowHammer mitigation techniques and evaluate them. 
The case study focuses on three aspects: 
\begin{itemize}
    \item \textbf{Functionality}: We confirm that our framework can reliably induce inter-VM RowHammer.
    \item \textbf{Security}: We show that the mitigation techniques can prevent inter-VM RowHammer in our tested cases.
    \item \textbf{Performance overhead}: We measure the VM performance under the mitigation techniques and compare it against the vanilla case with no mitigation applied.
\end{itemize}

\begin{table}[t]
\caption{Evaluation Environment and Parameters}
\label{table:environment}
\centering
\begin{tabular}{|l|l|}
\hline
                                & \multicolumn{1}{c|}{Configurations}                                                                                                                               \\ \hline
\multirow{3}{*}{gem5}           & version 24.1.0.0                                                                                                                                                  \\ \cline{2-2} 
                                & CPU: 3 cores OoO (Out of Order)                                                                                                                                   \\ \cline{2-2} 
                                & Cache Model: TwoLevelCacheHierarchy                                                                                                                               \\ \hline
\multirow{5}{*}{DRAMsim3}       & DRAM Model: DDR4\_4Gb\_x8\_2400                                                                                                                                               \\ \cline{2-2} 
                                & RowBufferPolicy: OpenPage                                                                                                                                         \\ \cline{2-2} 
                                & \multirow{3}{*}{\begin{tabular}[c]{@{}l@{}}1 Channel, 1 Rank, 4 BankGroup,\\ 2 Bank, 65536 rows, and 8192 columns,\\ 128 subarray groups (512 rows each)\end{tabular}} \\
                                &                                                                                                                                                                   \\
                                &                                                                                                                                                                   \\ \hline
Hammulator                      & HC\_first: 50K                                                                                                                                                    \\ \hline
\multirow{4}{*}{Software Stack} & Linux v6.1.0                                                                                                                                                      \\ \cline{2-2} 
                                & Bao demo                                                                                                                                                          \\ \cline{2-2} 
                                & U-Boot 2022.10                                                                                                                                                    \\ \cline{2-2} 
                                & ARM Trusted Firmware-A v2.9.0                                                                                                                                     \\ \hline
\end{tabular}
\end{table}

The parameters for our simulation environment are summarized in Table~\ref{table:environment}.
We configure our framework to simulate an out-of-order CPU with three cores and 4~GiB of DRAM.
This DRAM has 1 channel, 1 rank, 4 bank groups, 2 banks, 65536 rows, and 8192 columns.
The rows are further divided into 128 subarray groups (512 rows each) internally.
We set the \texttt{HC\_first} parameter of Hammulator to 50K, which means that a bitflip is probabilistically induced after 50,000 activations within a single refresh interval.

Each VM is given a contiguous memory region of 512~MiB.
The host machine is equipped with an Intel Core i5-12600KF with 64~GiB of memory and runs Ubuntu 22.04 LTS.

\subsection{Reproducing Mitigation Techniques}
\label{subsec:reproduced_mitigation_techniques}
To reproduce \textbf{Siloz}, we calculate PA ranges that are contained in different subarray groups by considering the DRAM address mapping specified.
We choose two contiguous PA ranges from them and assign them to the two VMs using the interface in our framework.
This ensures that the memory region of each VM is contained in a subarray group.

To reproduce \textbf{Citadel}, we calculate PA ranges that correspond to global rows by considering the DRAM address mapping specified.
We use the interface in our framework to specify a PA range among them as an unused region.
The VMs are assigned PA ranges that sandwich this unused region in the DRAM row space.
Note that this method does not reproduce all possible cases in Citadel
because our current framework can only assign a contiguous PA range to a VM.

\subsection{Setups}
\label{sec:framework_setup}
We describe the experimental setups for defining DRAM address mappings and inducing inter-VM RowHammer (i.e., evaluation steps 1 and 3 in Section~\ref{sec:proposed_framework_overview}).

{\bf Define DRAM Address Mappings}:
We use three representative DRAM address mappings from common categories analyzed in prior work~\cite{Pessl2016}.
A mapping is defined as a set of PA bits ($x_i$) to index DRAM coordinates.
Below, $f_{\text{foo}}$ represents the index for foo.
For example, $f_{\text{bank}} = x_0$ means that an access is served by bank $b \in \{0, 1\}$ if the least significant bit of the address is $b$. 
For all mappings, $f_{\text{column}}$ and $f_{\text{bankgroup}}$ are fixed to  $x_{12..0}$ and $x_{14,13}$, respectively.
There are no indices for channels and ranks because our DRAM only has one each.
The other indices are defined as follows.

\begin{itemize}
    \item \textbf{Simple mapping}: 
    This baseline configuration uses contiguous PA bit fields: $f_{\text{bank}} = x_{31}$ and $f_{\text{row}} = x_{30..15}$.
    \item \textbf{Bank XOR mapping}: 
    This mapping introduces XOR for bank indexing while keeping row indexing contiguous: $f_{\text{bank}} = x_{31} \oplus x_{6}$ and $f_{\text{row}} = x_{30..15}$.
    \item \textbf{Bank XOR and non-contiguous row mapping}: 
    This complex mapping uses XOR for bank indexing and non-contiguous bits for row indexing: $f_{\text{bank}} = x_{21} \oplus x_{6}$ and $f_{\text{row}} = x_{31..22,20..15}$.
    It is named non-contiguous row mapping because $x_{21}$ is missing in $f_{\text{row}}$.
\end{itemize}

{\bf Inducing Inter-VM RowHammer}: The attacker VM and the attacker program running on it are configured as follows.
We assume that the attacker knows the DRAM address mapping and which PA corresponds to rows adjacent to the victim VM.
The attacker VM runs Linux kernel 6.1.0 compiled with the \texttt{CONFIG\_STRICT\_DEVMEM} option disabled. 
This allows user-space applications within the attacker VM to access {\bf any} PA via \texttt{/dev/mem}.
The attacker program utilizes it and the cache flush instruction (\texttt{dc civac}) to repeatedly access the PA corresponding to an aggressor row adjacent to the memory region of the victim VM.
The hypervisor passes through this instruction to the underlying simulated hardware.
Although we use Linux for easy setup, the user of our framework can choose any method that runs as a guest OS (e.g., writing their own bare-metal program that induces RowHammer).

\subsection{Procedures and Results}
\subsubsection{Functionality}
\label{inter-vm-rowhammer-attack-simulation}

\begin{table}[t]
\caption{Wall-clock Time (in seconds, measured in the host) until the first bitflip is observed.}
\label{table:sim-time}
\centering
\begin{tabular}{|l|l|}
\hline
                & Wall-clock time (sec) \\ \hline
From boot       & 4277                          \\ \hline
From checkpoint & 123                           \\ \hline
\end{tabular}
\end{table}

To verify the framework's ability to induce inter-VM RowHammer, we execute the attacker program within a VM (attacker VM).
A checker program running in the other VM (victim VM) periodically reads data from a PA in its own memory region that is adjacent to the aggressor row.
Inter-VM RowHammer is considered successful when the checker program detects changes in the read data. 

The result of this experiment is twofold.
First, we confirmed that a bitflip was observed in the victim VM in all the address mappings.
This means that our framework can simulate inter-VM RowHammer scenarios in various DRAM address mappings.
Note that the bitflips are not caused by any bugs because they occurred in and only in the rows adjacent to the ones we hammer.
Second, we measured the wall-clock time (in the host) that it tool to observe the first bitflip, shown in Table~\ref{table:sim-time}.
We tested with the XOR address mapping as a representative case.
The label ``from boot'' indicates that the simulation was executed from the beginning,
while ``from checkpoint'' indicates that it was executed from a checkpoint where the VMs have finished their boot processes.
In the former case, observing the first bitflip took approximately 1.16 hours, while it only took around 2 minutes in the latter.
Due to this large speedup, we use the same methodology (starting a simulation from a checkpoint) in the subsequent experiments.

\subsubsection{Security}
\begin{table}[t]
\caption{Security evaluation across different DRAM address mappings. The symbol \Checkmark indicates that inter-VM RowHammer was mitigated.}
\label{table:security-results}
\centering
\begin{tabular}{|c|c|c|c|}
\hline
 & Simple & XOR & XOR and Non-Contiguous Row \\ \hline
Siloz   & \Checkmark & \Checkmark & \Checkmark \\ \hline
Citadel & \Checkmark &  \Checkmark &  \Checkmark \\ \hline
\end{tabular}
\end{table}

To evaluate Siloz and Citadel in our framework, we first change the memory allocation of the two VMs in accordance with the allocation policy of each mitigation technique (e.g., assigning different subarray groups to them).
After that, we launch the VMs normally and try to induce inter-VM RowHammer as we do in the functionality experiment.

Table~\ref{table:security-results} shows the results.
We observed that both Siloz and Citadel mitigated inter-VM RowHammer in any DRAM address mapping tested.
In the Siloz cases, we confirmed that bitflips occurred within the subarrays assigned to the attacker VM.
In the Citadel cases, we detected bitflips within the guard rows inserted between the VMs.
It is important to note that we only claim that these mitigation techniques prevent inter-VM RowHammer in our tested configurations (e.g., address mappings, VM region sizes).
Our framework makes it possible to conduct this kind of evaluation on various configurations.

\subsubsection{Performance overhead}
To quantify the performance overhead, we measure the elapsed times of (1) the boot process of the VMs, and (2) a matrix-vector multiplication program executed on the VMs.
The elapsed times are measured in the unit of simulated seconds and acquired from the stat file of gem5.
The measurement of the VM boot process starts when the hypervisor has finished its initialization and jumps to the bootloader of the first VM (out of the two),
and finishes when both VMs have done booting Linux.
The measurement for the matrix-vector multiplication does not include the VM boot time.

\begin{figure}
    \centering
    \includegraphics[width=1\columnwidth]{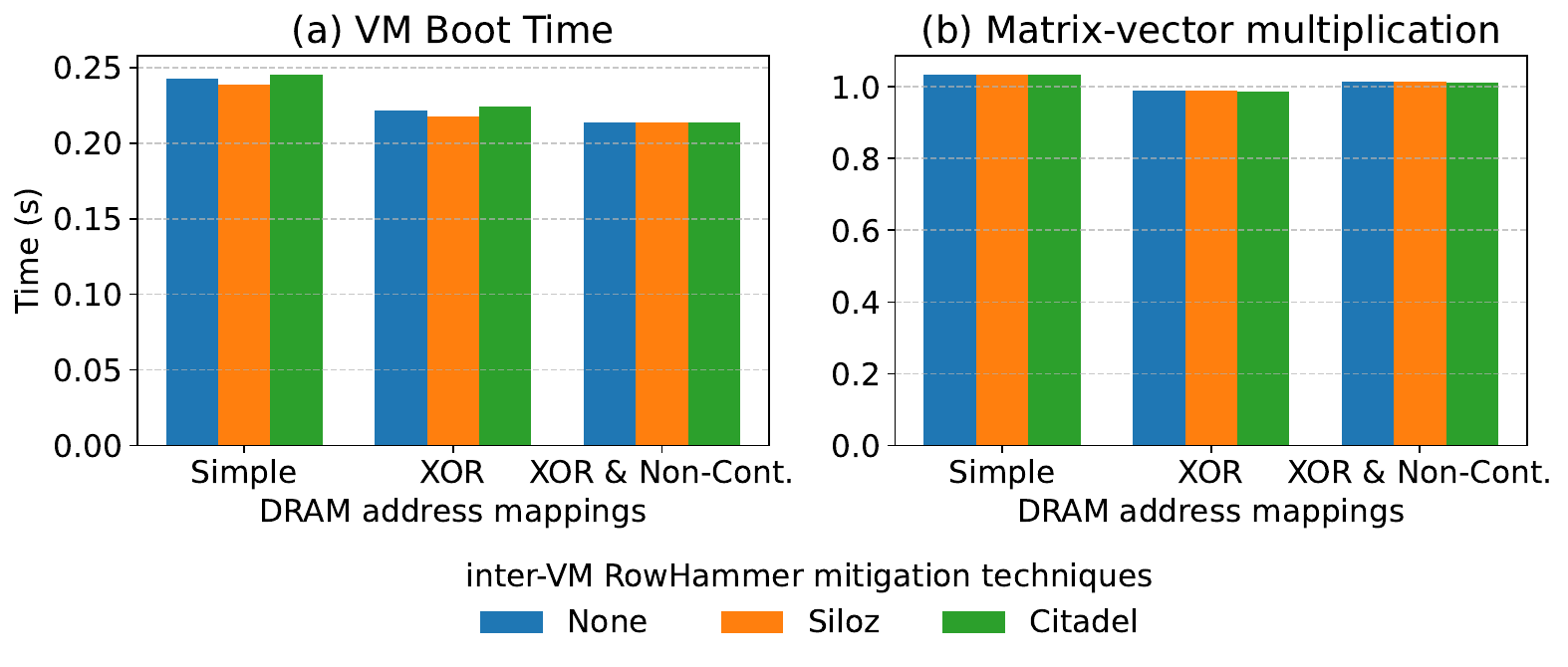}
    \caption{(a) VM boot time and (b) Execution time of matrix-vector multiplication (matrix size: 256 MiB)}
    \label{fig:overhead}
\end{figure}

Fig.~\ref{fig:overhead}~(a) and Fig.~\ref{fig:overhead}~(b) show the results for the boot time and matrix-vector multiplication, respectively.
The bars labeled as ``None'' are  the results when no mitigation technique is applied.
%We make three observations from the results.
Three observations are made from the results.
First, the results differed in VM boot time but did not in matrix-vector multiplication.
Second, the XOR-based mappings yielded faster boot times compared to Simple.
Third, Siloz was slightly faster than None in VM boot time while Citadel was slower.
We hypothesize that these observations stem from different DRAM performance due to different data addresses.
The Linux kernel is placed statically by the boatloader, thus its alignment is affected by the starting address of the VM memory region,
which in turn is affected by the mitigation technique (e.g., guard rows in Citadel).
On the other hand, the addresses of the matrix and the vector are decided by libc (we use \texttt{new}).
This could conceal the different alignments of the starting addresses of the VM memory regions.
Although we do not further analyze the phenomena in this paper,
the important point is that our framework enables this kind of analyses through simulation on various DRAM address mappings.

\section{Discussion}
\label{Discussion}

\subsection{Related Work}
DRAM address mappings can be configured to some extent in limited hardware~\cite{AMDBIOS} 
and prior work utilizes this functionality by writing a designated UEFI shell scripts~\cite{Hillenbrand2017}. 
However, this method only provides an interface to toggle channel and bank interleaving, but not to define an arbitrary address mapping unlike our framework provides.

Existing simulators such as Hammulator~\cite{Thomas2023} and Hemmersim~\cite{Kaustav2023} can model RowHammer effects inside DRAM.
They could be more useful than our framework to evaluate hardware-based mitigation techniques such as MINT~\cite{Qureshi2024}.
However, they cannot directly reproduce software-based mitigation techniques as they only model the hardware side.
%Unlike these previous work, our research focuses on the inter-VM RowHammer scenarios.

There are several studies on running hypervisors on micro-architecture simulators~\cite{Peter2021, George2024}.
Peter et al.~\cite{Peter2021} achieved full-system RISC-V simulation in gem5. 
They demonstrated Linux booting on a hypervisor in the M-mode defined in RISC-V.
George-Marios et al.~\cite{George2024} implemented the RISC-V hardware-assisted virtualization extensions in gem5. 

\subsection{Internal DRAM Behaviors}
Another hurdle of RowHammer attacks besides DRAM address mappings for both attackers and defenders is the internal behaviors of DRAM chips.
They include the row coupling effect~\cite{Hwayong2024} and internal address mappings~\cite{Hou2016}.
For example, rows specified by the memory controller can be remapped internally in the DRAP chip to avoid faulty rows.

Considering internal DRAM behaviors is the responsibility of DRAM simulators (e.g., DRAMSim3) and the user of our framework, but not the framework itself.
Once a DRAM simulator supports simulating these behaviors, our framework can be extended accordingly so that the user can specify them from our framework to the underlying simulator.

\subsection{Limitations}
{\bf RowHammer between VM and Hypervisor}:
Siloz and Citadel consider RowHammer-induced bitflips targeting the hypervisor. 
In this paper, we focus specifically on inter-VM RowHammer. 
Extending our framework to address attacks targeting the hypervisor is left for future work.

{\bf Inter-VM RowHammer via SLAT}:
Hardware virtualization support typically includes second-level address translation (SLAT).
SLAT translates guest physical addresses (GPAs) to host physical addresses (HPAs). 
Bitflips induced by RowHammer within these structures could corrupt this mapping.
This corruption allows a VM to access arbitrary HPAs and compromise isolation between VMs~\cite{Chen2025}.
Our current framework employs a straight GPA-to-HPA mapping that makes complex SLAT lookups unnecessary and thus not performed.
Therefore, attacks targeting SLAT mechanisms currently lie outside the scope of our evaluation.

{\bf Non-Contiguous Physical Address Allocation}:
Our framework currently only supports allocating a single contiguous host PA range per VM.
This forbids our framework from reproducing some defense scenarios in Citadel.
To support non-contiguous host PA ranges, we need to extend Bao (or use other hypervisor)
so that the SLAT is fully configured to support GPA-to-HPA mappings other than straight.

\section{Conclusion}
\label{Conclusion}
This paper proposed a simulation-based framework to evaluate inter-VM RowHammer mitigation techniques under various DRAM address mappings.
By reproducing mitigation techniques in a lightweight hypervisor on top of an architecture simulator,
we enable security and performance evaluation of them without intense reverse-engineering to acquire address mappings from real hardware.
Our case study showed that it can simulate inter-VM RowHammer as well as enable evaluation of two existing inter-VM mitigation techniques.

\section*{Acknowledgements}
This work was supported by JST, PRESTO Grant Number JPMJPR22P1, Japan.
We thank the anonymous reviewers for their valuable feedback to improve this paper.

\bibliographystyle{IEEEtran}
\bibliography{main}

\end{document}